\documentclass[twocolumn,english,aps,prb,showpacs,superscriptaddress]{revtex4-1}
\usepackage[latin9]{inputenc}
\usepackage{amsmath}
\usepackage{amssymb}
\usepackage{graphicx}
\usepackage{esint}
\usepackage{physics}
\usepackage{enumitem}
\usepackage{braket}

\makeatletter

\usepackage{physics}
\usepackage{color}

\makeatother

\usepackage{babel}

\begin{document}

\title{On the topological immunity of corner states in two-dimensional crystalline insulators}

\author{Guido van Miert}
\affiliation{Dipartimento di Fisica ``E. R. Caianiello'',
Universit\`a di Salerno I-84084 Fisciano (Salerno), Italy }

\author{Carmine Ortix}

\affiliation{Dipartimento di Fisica ``E. R. Caianiello'',
Universit\`a di Salerno I-84084 Fisciano (Salerno), Italy }
\affiliation{Institute for Theoretical Physics, Center for Extreme Matter and
Emergent Phenomena,\\ Utrecht University, Princetonplein 5, 3584
CC Utrecht, the Netherlands}

\date{\today}

\begin{abstract}
A higher-order topological insulator (HOTI) in two dimensions is an insulator without metallic edge states but with robust zero-dimensional topological boundary modes localized at its corners. 
Yet, these corner modes 
do not carry a clear signature of their topology as they 
lack the anomalous nature of helical or chiral boundary states.
Here, we demonstrate using immunity tests  that the corner modes found in the breathing kagome lattice represent a prime example of a mistaken identity.
Contrary to previous theoretical and experimental claims, we show that these corner modes are inherently fragile:
 the 
kagome lattice
does not realize a higher-order topological insulator. 
We support this finding by introducing a 
criterion based on a corner charge-mode correspondence for the presence of topological midgap corner modes in 
$n$-fold rotational symmetric chiral insulators that explicitly precludes
the existence of a HOTI protected by a threefold rotational symmetry. 
\end{abstract}

\maketitle

\section{Introduction}
Topology and geometry find many applications in contemporary physics, ranging from anomalies in gauge theories to string theory. 
In condensed matter physics, topology is 
used to classify defects in nematic crystals, characterize magnetic skyrmions, and predict the presence or absence of (anomalous) metallic states at the boundaries of insulators and superconductors~\cite{has10,qi11}. For the latter, the topological nature of the boundary modes,
be they 
point-like zero modes~\cite{fu08,lut10,mou12}, one-dimensional chiral~\cite{tho82,hal82}
and
helical states~\cite{kan05,kan05b,fu06,ber06,mol07}, or two-dimensional surface Dirac cones~\cite{moo07,fu07b,fu07,zha09,ras13,bru11}, 
resides in their robustness. One
can only get rid of these states by a bulk band-gap closing and reopening
or
by breaking the protecting symmetry, 
which can be either an internal or a crystalline symmetry. 
For example, in two-dimensional topological insulators~\cite{kan05,kan05b,fu06,ber06,mol07} one can gap out the helical 
edge
states 
by 
introducing
a Zeeman term that explicitly breaks the protecting time-reversal symmetry. 
Similarly, one can move the 
end
states 
of a Su-Schrieffer-Heeger~\cite{su79}
chain away from zero energy by breaking the 
chiral (sublattice)
symmetry at the edges and/or in the bulk. 

Recent theories exploiting the protecting role of crystalline symmetries have  led to the discovery of the so-called higher-order topological insulators~\cite{fan17,ben17,ben17-2,sch18,sch18b,gei18,pet18,ser18,imh18,kha18b,lan17,sit12,son17,eza18b,eza18c,mie18,koo18} (HOTI): states of matter characterized by the presence of
topologically protected modes living at the $D-n$ dimensional boundary of a $D$ dimensional insulator, 
with $n>1$ denoting the order. 
Thus, a two-dimensional second-order topological insulators features point-like corner modes, while a three-dimensional second-order topological insulator features helical or chiral edge modes along the one-dimensional edges. 
The prediction of higher-order topological insulators has triggered an enormous interest in scanning material structures and engineering metamaterials, {\it e.g.} electric circuits~\cite{imh18},
exhibiting topological corner or hinge modes. However, in identifying higher-order topological insulating phases different complications arise. First, as in conventional first-order topological insulators, the system can display ordinary in-gap boundary states~\cite{zak85}
that are not the prime physical consequence of a non-trivial bulk topology.
Second, $D-2$ boundary modes of a $D$-dimensional system can be a manifestation of the crystalline topology of the $D-1$ edge rather than of the bulk: the corresponding insulating phases have been recently dubbed as boundary-obstructed topological phases~\cite{kha19} and do not represent genuine (higher-order) topological phases.
These complications 
are particularly severe in 
second-order higher-order topological insulators in two dimensions, 
since 
the corresponding zero-dimensional topological boundary modes fail to possess the anomalous nature chracteristic of, for instance, one-dimensional chiral modes 
or zero-dimensional Majorana corner modes in second-order topological superconductors.
Singling out a proper second-order topological insulator in two dimensions thus represents a task of exceptional difficulty. In this paper, we prove this assertion  by showing that one of the first model system suggested to be a second-order topological insulator in two-dimension~\cite{eza18,kem19,xue19,ni19,li20} --  the breathing kagome lattice model --   does not host any higher-order topological phase. The corner modes experimentally found in this system~\cite{kem19,xue19,li20} are neither engendered from a second-order bulk topology nor from the edge topology characterizing boundary-obstructed topological phases. They instead are an example of conventional corner modes, with the exact same nature of the edge modes generally appearing in one-dimensional insulating chains~\cite{mie17}. 
We contrast the fragility of the corner modes in the kagome lattice with the robustness of the corner modes in chiral-symmetric insulators, which possess the full immunity of the topological corner modes of a HOTI when an additional even-fold rotational symmetry is present. We also formulate a one-to-one correspondence between fractional corner charges~\cite{mie18,ben19}and corner modes, which 
predicts the presence or absence of topological zero modes in chiral symmetric insulators with a ${\mathcal C}_n$-rotational symmetry, and explicitly precludes the appearance of a HOTI phase protected by a threefold rotational symmetry.

\begin{figure}
\includegraphics[width=1\columnwidth]{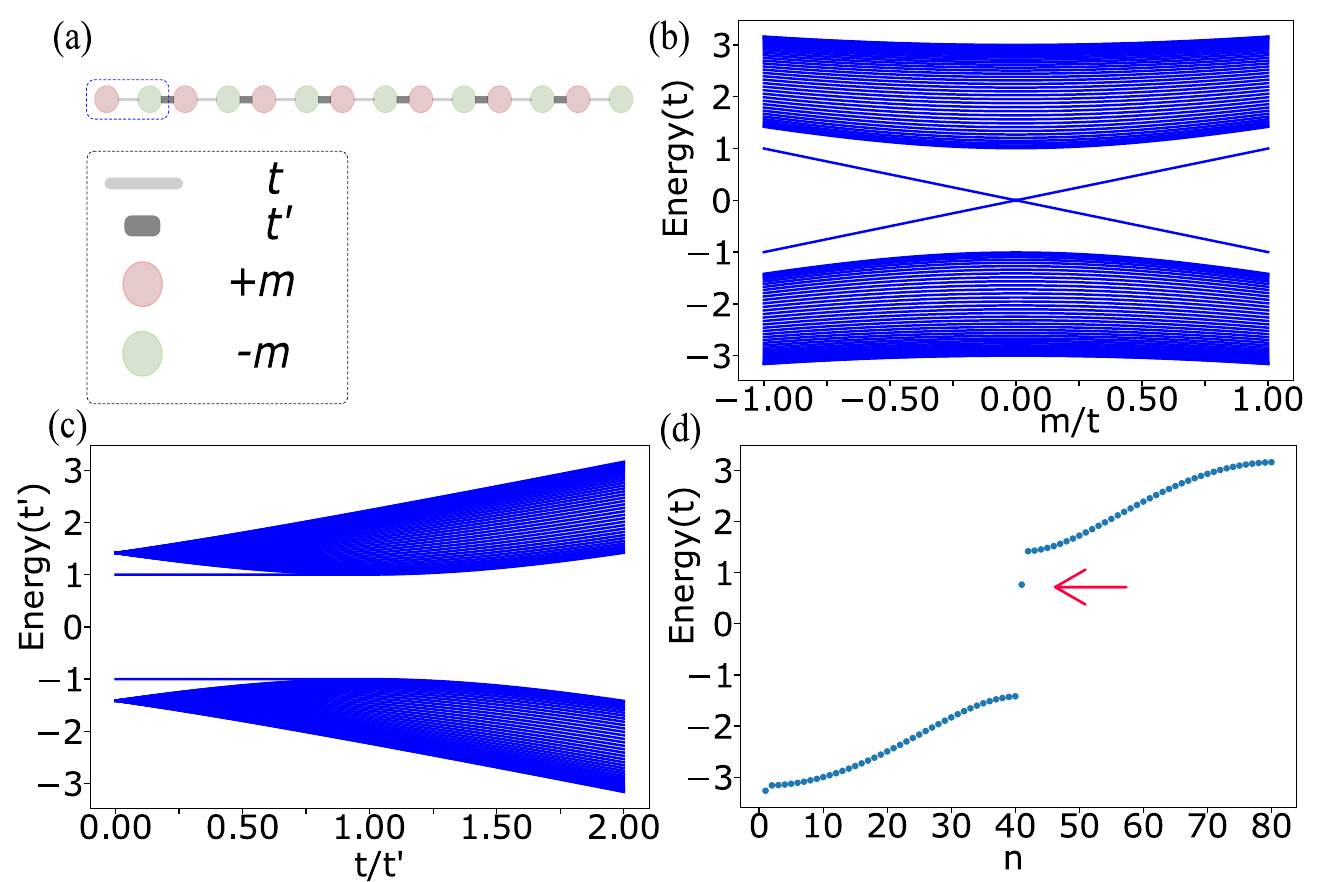}
\caption{{\bf Non-topological boundary modes in atomic binary chains.} (a) Sketch of a Rice-Mele atomic chain	 containing $7$ unit-cells. (b) Energy spectrum of a finite Rice-Mele chain containing $40$ unit-cells as obtained by varying $m$ (measured in units of $t$). The intercell hopping amplitude $t' = 2 t$. Energies have been measured in units of $t$.  (c) Energy 
spectrum of a finite Rice-Mele atomic chain containing $40$ unit-cells as a function of the hopping amplitude $t$ (measured in units of $t'$). The on-site energy parameter $m = t'$. Energies have been measured in units of $t^{\prime}$. 
 (d) Energy levels of a finite Rice-Mele atomic chain with $40$ unit-cells where the we have changed the on-site energy on the left edge site from $+m$ to $-m$. This lead to the emergence of an in-gap state. We have used the following parameters: $t' = 0.5 t $, and $m=0.5 t$. Energies have been measured in units of $t$. 
 \label{fig:RMchain}}
\end{figure}

That boundary in-gap modes can be encountered in both a topological and a non-topological host system  is immediately shown by analazying the simple example of one-dimensional band insulators, which we recall here to simplify the discussion of our results below.
Let us consider specifically a paradigmatic
minimal model for a one-dimensional band insulator: 
the Rice-Mele model~\cite{ric82}.
It is
schematically 
shown
in Fig.~\ref{fig:RMchain}(a). 
The electrons living on the 
red --  sublattice A (green -- sublattice B) sites
experience an on-site energy $+m$ ($-m$),
can hop within the unit-cell 
with hopping amplitude $t$,
and between adjacent unit-cells 
with a hopping amplitude $t^{\prime}$.
Within the bulk energy gap 
and when
$|t/t'|<1$ we find that the left edge hosts a state at energy $+m$, whereas the right edge hosts a state at energy $-m$
[c.f. Fig.~\ref{fig:RMchain}(b)].
On the contrary,
when $|t/t'|>1$ the system fails to exhibit any boundary state
[c.f. Fig.~\ref{fig:RMchain}(c)].
 Hence, the Rice-Mele model 
exhibits boundary states in half of the parameter space,   
assuming the termination shown in Fig.~\ref{fig:RMchain}(a) is chosen.
In order to establish whether these boundary states are topological in nature, we resort to the main characteristic of topological boundary states: their robustness against smooth perturbations. We notice the following:

\begin{enumerate}[label=(\roman*)]
\item The energies of the in-gap boundary states change either introducing smooth deformations of the bulk Hamiltonian or applying edge-specific perturbations. An example of this is illustrated in Fig.~\ref{fig:RMchain}(b). 
\item The boundary states
can completely dissolve into the bulk bands 
without an accompanying bulk-band gap closing, as shown in Fig.~\ref{fig:RMchain}(c). 
In the Rice-Mele model this occurs for $|t/t'|=1$.
\item A tailor-made edge potential can even lead to the creation of boundary states out of the bulk bands in the parameter region $|t/t'|>1$, as explicitly shown in Fig.~\ref{fig:RMchain}(d).
\end{enumerate}
The boundary states
of the Rice-Mele model fail to exhibit any kind of robustness, and can be therefore qualified as ordinary, {\it i.e.} non-topological boundary states. 

This, however, is not yet the end of the story. In fact, in one-dimensional insulating models it is possible to exploit the effect of the internal particle-hole and chiral symmetries.
Their existence  implies that the model Hamiltonian anticommutes with an (anti)unitary operator  that squares to $1$ ($\pm 1$).
The Bogoliubov-de Gennes (BdG) Hamiltonian describing a superconductor is, by its very definition, particle-hole symmetric. 
However, both the particle-hole
and the chiral symmetry also play a 
role outside the superconducting realm.
They can in fact arise as approximate symmetries of 
the effective model Hamiltonian describing an insulator. As an example, in the Rice-Mele chain we may set $m=0$, in which case the Hamiltonian anti-commutes with $\sigma_z$. The resulting model is 
the well-known
Su-Schrieffer-Heeger (SSH) chain.

Using the results above, we find that the SSH chain displays
a left and a right 
boundary 
state at zero-energy if and only if $|t/t'|<1$.
Importantly, these boundary states now represent truly topological boundary states.
Obviously, it would amount to cherry-picking 
if these states would be dubbed as topological
only because they are at zero energy. The rationale for 
the above is instead
based on the following fact: The boundary modes cannot be moved away from zero energy using chiral symmetry preserving bulk or edge perturbations, 
as long as these do not close the bulk band gap. For example, long-range hoppings between the two sublattices do not move the boundary states since the chiral symmetry is preserved when these processes are included. Hence, the boundary states of the SSH chain are robust zero-energy modes that are protected by the internal chiral symmetry. 

\begin{figure}
\includegraphics[width=1\columnwidth]{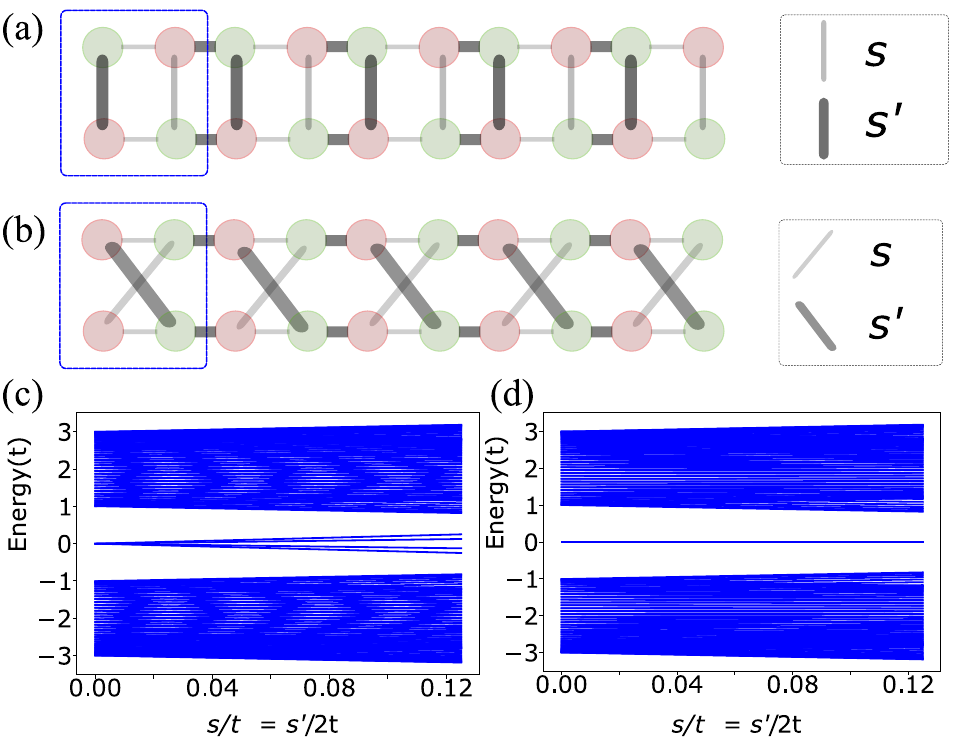}
\caption{{\bf Boundary modes of chiral one-dimensional insulator verify ${\mathbb Z}$ group law.} (a)-(b) Sketch of a two-leg atomic chain obtained by hybridizing two SSH atomic chain in different chiral symmetry preserving manners. The hybridization hopping parameters are denoted as $s$ and $s^{\prime}$. 
schematic depiction of hybridized SSH chains. (c) -(d) Corresponding evolution of the spectrum as a function of $s = s'/2$ (measured in units of $t$) . For both cases we have taken a chain of $40$ unit-cells, with $t'=2 t$. Energies have been measured in units of $t$. 
\label{fig:SSHchainhybr}}
\end{figure}

Moreover, and as shown in Appendix A,
we can topologically discriminate between systems with an odd and even number of left (right) edge states pinned at zero energy.
Therefore,
 one-dimensional chiral-symmetric insulators -- but the argument above holds true also for particle-hole symmetric insulators belonging to the class D of the Altland-Zirnbauer table~\cite{alt97,kit09} --
are characterized by the group $\mathbb{Z}_2 = \{0,1\}$, with the identity element $0$ corresponding to chains featuring an even number of left edge states, and the element $1$ corresponding to chains featuring an odd number of left edge states. Physically, the $\mathbb{Z}_2$ group law translates into the fact that if one combines (i.e. hybridizes in a symmetry-preserving way) a chain with an even number of left edge states and a  chain with an odd number of left edge states, 
the combined two-leg atomic chain possesses
an odd number of left edge states.
On the contrary, hybridizing two chains with both an even or odd number of left edge states results in a two-leg atomic
chain with an even number of left edge states. 
In Fig.~\ref{fig:SSHchainhybr} we illustrate this by considering a two-leg atomic chain consisting of two SSH chains, each of which featuring a single left edge state, 
hybridized in two different ways. In the first example, shown in Fig.~\ref{fig:SSHchainhybr}(a) we have hybridized one SSH chain that terminates with an A site, with another SSH chain that terminates with a B site. The result of this hybridization, is that the two left edge states also hybridize, 
and move away from zero energy, as explicitly shown in Fig.~\ref{fig:SSHchainhybr}(c).
This is simply a manifestation of level repulsion. However, the two-leg atomic chain shown in Fig.~\ref{fig:SSHchainhybr}(b) displays a 
more interesting behavior. There we have hybridized two SSH chains that both terminate with the same sublattice. Despite the hybridization, we find that the two edge states do not move away from zero
energy [c.f. Fig.~\ref{fig:SSHchainhybr}(d)].
Even though these two examples provide purely anecdotal evidence, 
they do  suggest that the physics of zero-energy edge states is not fully captured by a $\mathbb{Z}_2$-invariant.

This is indeed the case for 
chiral-symmetric insulators. 
In these systems, the isolated zero energy modes must be eigenstates of the unitary chiral symmetry, and can be consequently characterized by their chiral charge. For the specific example of the SSH atomic chain, this also implies that an isolated zero mode is either fully localized on the A-sublattice or on the B-sublattice. 
Denoting with $\ket{\Psi}$ the zero energy end state of a SSH chain and with $\chi= \bra{\Psi} \sigma_z \ket{\Psi}$ the corresponding chiral charge, we have that if $\chi=1$ the zero-energy mode will be fully localized on the sublattice A, whereas if $\chi=-1$ the zero energy mode will be fully localized on the sublattice B. The perfect localization of the end states clearly implies that having at hand two zero-energy states localized on the same sublattice, and thus with same chiral charge, impedes any level repulsion as this would necessarily break the chiral symmetry. On the contrary, a pair of zero-energy modes localized on different sublattices can be opportunely coupled and moved away from zero energy, in perfect agreement with the features of the two-leg atomic chain shown in Fig.~\ref{fig:SSHchainhybr}(c),(d). 
This proves that
the physics of zero energy states in chiral-symmetric insulators is encoded in the $\mathbb{Z}$-valued chiral charge: 
\begin{align*}
\chi_L = \sum_j \langle\Psi_j|\sigma_z|\Psi_j\rangle \in \mathbb{Z},
\end{align*}
where the sum runs over all left edge states $|\Psi_j\rangle$. Note, however, that the relation between the chiral charge $\chi_L$ and the number of left edge states at zero energy does not represent a one-to-one correspondence.
With a zero chiral charge, a pair of zero energy boundary states can be still encountered. This thus implies that the
absolute value of the left chiral edge charge defines a lower bound for the total number of zero-energy states, and is equal to the number of edge states {\it modulo} $2$. 
Hybridizing
two insulators with chiral charges $n_1$ and $n_2$, respectively, we will end up with a chain that features at least $|n_1+n_2|$ edge states at zero energy. 

To summarize, boundary states
are a generic feature of one-dimensional band insulators. However, in the absence of any symmetry 
these states are not robust, for instance they can dissolve into the bulk bands, or even created as a result of tailor-made edge perturbations.
Instead, if one considers systems with a particle-hole or chiral symmetry, we find that the parity of the number of left and right edge states is robust against symmetry allowed perturbations. Moreover, for systems with a chiral symmetry we find that there is a chiral charge associated to each edge, which represents a $\mathbb{Z}$ number and is a lower bound for the number of edge states.

\section{Results}

\subsection*{Corner modes in the breathing kagome lattice}
\begin{figure}
\includegraphics[width=1\columnwidth]{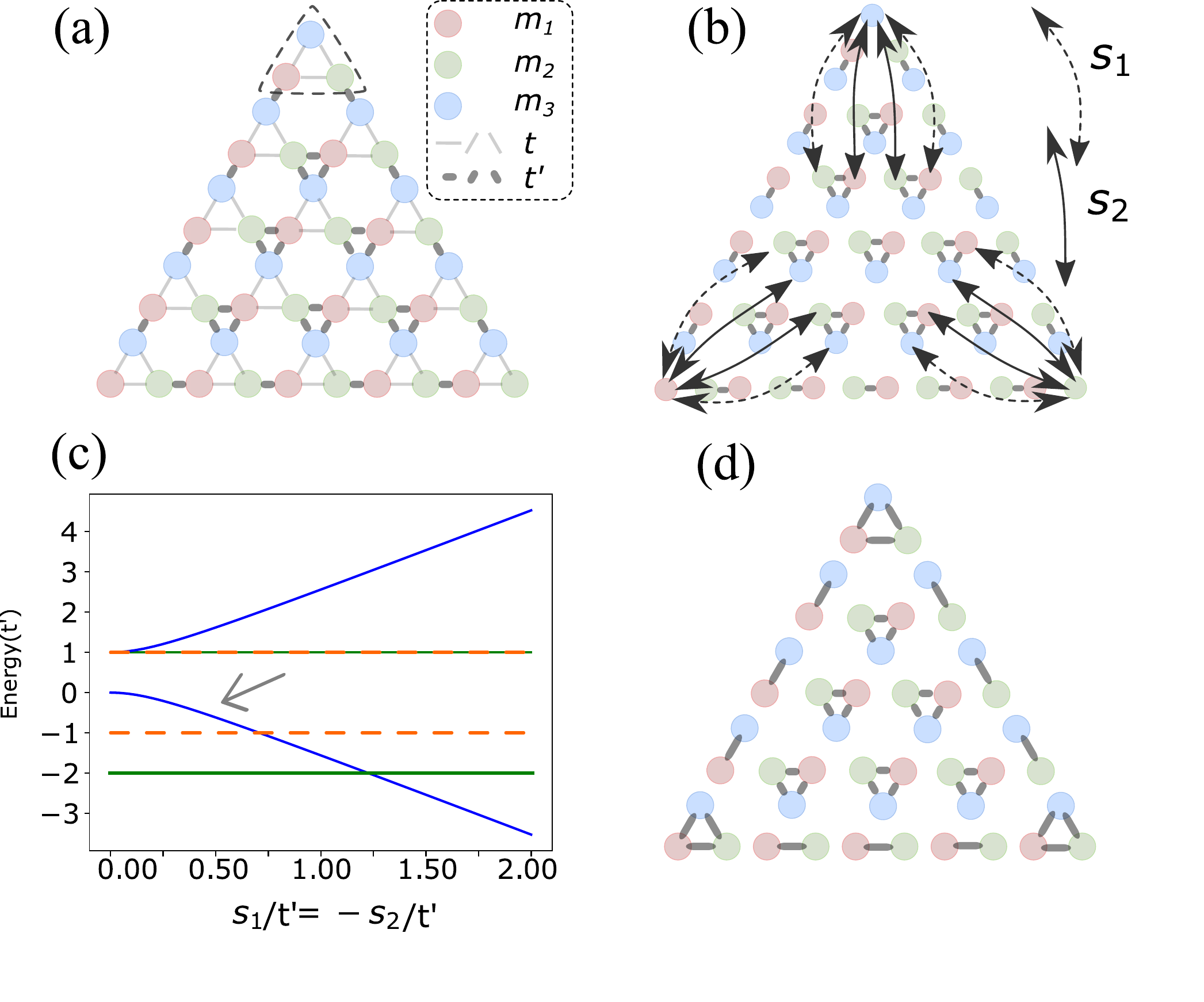}
\caption{{\bf Corner modes of the breathing Kagome lattice fail immunity tests.} (a) Schematic depiction of the breathing kagome lattice. (b) Local perturbation applied at the corners of the kagome lattice, respecting the three-fold rotational symmetry, the two-fold mirror symmetries, and the generalized chiral symmetry. (c) Evolution of the spectrum of the kagome lattice as a function of the perturbation stength $s_1 = - s_2$ (measured in units of $t^{\prime}$).
Here, we have used $t = 0$, and energies have been measured in units of $t^{\prime}$. 
The gray arrow indicates the evolution of the corner modes.The bulk states are shown in green, the edge states in orange and the states localized at the corner in blue. The shown spectrum applies to lattices with four or more  unit-cells along each edge. (d) Edge perturbation that interchanges the values of the intra- and inter-cell hopping amplitudes $t$ and $t^{\prime}$ along the lattice boundaries.  \label{fig:kagome1}}
\end{figure}

The preceding discussion on the physical properties of end states in one-dimensional insulators can be applied
in an analogous fashion to corner states in two-dimensional insulators. 
This can be nicely illustrated using the breathing kagome lattice [c.f. Fig.~\ref{fig:kagome1}(a)], which can
be thought of as the two-dimensional cousin of the Rice-Mele 
atomic chain. For simplicity, we 
will  assume 
in the remainder
that all intra-unit cell hopping 
amplitudes
have an equal magnitude $t$. We will also make the same assumption for the inter-unit cell hopping amplitudes (magnitude $t^{\prime}$).
In addition to the inter- and intra-unit cell hopping parameters, 
we will first allow for different on-site energies in the three sublattices, which we denote with
$m_1$, $m_2$, and $m_3$, respectively. 
We note that in this situation the breathing kagome lattice is in the wallpaper group $p1$, {\it i.e.} there are no crystalline symmetries other than the in-plane translations.

Following Ref.~\cite{kun18} we find that when considering the system in a open-disk geometry,
the lower left, lower right, and upper corner host corner states at energies $m_1$, $m_2$, and $m_3$ in half the parameter space, {\it i.e.} for $|t/t'|<1$.
This condition being very similar to the one we encountered in the Rice-Mele atomic chain, suggests that also the corner modes of the breathing kagome lattice correspond to conventional in-gap bound states. And indeed these modes dissolve into the bulk by changing the ratio $|t/t^{\prime}|$, see, e.g., Ref.~\cite{kun18}. 
Next, let us consider the situation in which the on-site energies of the three sublattices are constrained to be equal, {\it i.e.} $m_1 \equiv m_2 \equiv m_3 \equiv 0$. 
In this situation, the lattice is in the wallpaper group $p3m1$, which is generated by the point group ${\mathcal C}_{3v}$ with the addition of translations.
The fact that the corner modes now reside at zero energy in an extended region of the parameter space could suggest that the corner modes 
represent in this case genuine topological boundary modes. Contrary to the SSH atomic chain, however, the kagome lattice model does not possess an internal chiral symmetry that can protect the existence of corner modes pinned at zero energy. This follows from the very simple fact that  
a chiral-symmetric insulator is incompatible with an odd number of sublattices.
Nevertheless, recent studies have suggested that other symmetries could protect the existence of the zero-energy corner modes, qualifying them as topological corner modes and consequently the prime physical consequence of a higher-order non-trivial bulk topology. 
Specifically, in Ref.~\cite{eza18}  it has been suggested 
that the corner modes are protected by the combination of 
the three-fold rotation symmetry and the mirror symmetry of the point group ${\mathcal C}_{3v}$.
Instead, in Refs.~\cite{kem19,ni19,xue19}
a generalized chiral symmetry has been defined in order to prove the topological nature of the corner modes. Such a generalized chiral symmetry is equivalent to requiring that  the 
model Hamiltonian may only be perturbed by hopping processes 
between different sublattices. 

We now show that the opposite is true, and that
the corner modes found in the breathing kagome lattice are 
nothing but conventional boundary modes even for $m_{1,2,3} \equiv 0$. 
To simplify our discussion below, we will consider the flat-band limit of $t=0$. This limit has the main advantage of removing all system size dependence in our analysis.
As mentioned in the preceding section, the defining characteristic of any topological boundary mode is its immunity against perturbations that do not close the insulating band gap and preserve the protecting symmetries. In the context of 
two-dimensional insulators (without metallic edge states) in general, and of
the breathing kagome lattice in particular, this would mean that the corner modes have to remain pinned at zero energy upon perturbing the two-dimensional bulk, the one-dimensional edges, or the zero-dimensional corners. To see whether the breathing kagome lattice 
possesses boundary modes with such robustness,
we have considered the effect of applying a local perturbation at the three corners. Specifically, we have introduced long-range hopping processes, with amplitudes $s_1$ and $s_2$,
at the three corners, as schematically shown
in Fig. 3(b). Note that the corner perturbation 
fulfills all the symmetry constraints: the three-fold rotational symmetry, 
the mirror symmetries, and the generalized chiral symmetry are all preserved. 
Let us now consider the evolution of the corner state energy as we adiabatically 
  switch on the perturbation. In Fig. 3(c), we have plotted the evolution of the spectrum assuming, as mentioned above,
   the intracell hopping amplitude $t \equiv 0$ and the corner perturbation hopping amplitudes satisfy $s_1 \equiv -s_2$.
  We immediately find that the corner modes do not remain pinned at zero energy. In fact, we find that upon increasing the perturbation strength the energy of corner mode even crosses the edge and bulk valence bands at energies 
 $-t^{\prime}$ and $-2 t^{\prime}$, respectively. This demonstrates that the corner modes in the breathing kagome lattice do not possess any topological robustness. 
 Their presence or absence is not in a one-to-one correspondence with a topological invariant. Instead they simply constitute ordinary boundary states, as do the ones occurring when the on-site energies $m_{1,2,3}$
   are different from zero.
   
Even though the local corner perturbation considered above includes longer-range hoppings, we emphasize that the corner modes of the breathing kagome lattice are also unstable against perturbations that  involve short-range processes only. We illustrate this in Fig.~\ref{fig:kagome1}(d), where we consider an edge perturbation that interchanges the values of the intra- and inter-cell hopping amplitudes $t$ and $t^{\prime}$ along the lattice boundaries alone. Considering again for simplicity the limit $t=0$, we can readily infer that the system will fail to exhibit any in-gap corner mode, since the lattice is composed of disconnected dimers 
 whose bonding and antibonding states are at energies $\pm t^{\prime}$ and disconnected triangles with energies $-2 t^{\prime}$ and  $t^{\prime}$. 
 
 We finally wish to point out that the example of the breathing kagome lattice demonstrates that topological corner modes of insulators represent the exception rather than the rule. Therefore, in the absence of any mathematical or physical strong motivation supporting the topological nature of the corner modes, explicit immunity checks should be performed. In particular, these immunity checks should especially involve perturbations that respect all the putative protecting symmetries. In this respect, we note that Ref.~\cite{kem19} for instance has analyzed only the reaction of the kagome lattice corner modes under perturbations that explicitly broke the protecting symmetries. Furthermore, we emphasize that, as demonstrated above, immunity checks can be simply performed starting out from a flat-band, i.e. atomic, limit. Nevertheless, also perturbative approaches can be employed to shine light on the (non)-topological nature of the corner modes. Consider for instance, a zero energy corner mode $\ket{\chi}$ disturbed by a generic perturbation $\Delta V$. For the corner state to be topological, the energy of the corner mode should be pinned to zero to any order in $\Delta V$. Whereas different putative protecting symmetries can enforce the first-order perturbative correction to vanish, there is  {\it a priori} no symmetry able to guarantee the absence of a coupling between the corner state $\ket{\chi}$ and the edge or bulk states $\ket{\Psi}$. As a result, the second order correction to the corner state energy $\sum_{\Psi\neq\chi}\frac{\lvert\langle\Psi|\Delta V|\chi\rangle\rvert^2}{-E_{\Psi}}$ will generically differ from zero, except when an actual chiral symmetry is present. In fact, the latter guarantees that the finite contribution to the second-order correction coming from a state $\Psi$ is immediately cancelled out by the contribution due to its chiral partner $\widetilde{\Psi}$ with energy $E_{\widetilde{\Psi}}=-E_{\Psi}$. This further demonstrates the non-topological nature of the corner modes encountered in the breathing kagome lattice.

\subsection*{Corner modes in chiral-symmetric insulators}

Having established with a concrete microscopic model that the corner modes in the kagome lattice simply corresponds to fragile ordinary boundary modes, we next introduce a chiral-symmetric insulator featuring robust corner modes. This will also allow us to discuss the different nature and degree of protection 
provided by a non-trivial edge topology as compared to the bulk topology of a higher-order topological insulator. 

\begin{figure}
\includegraphics[width=1\columnwidth]{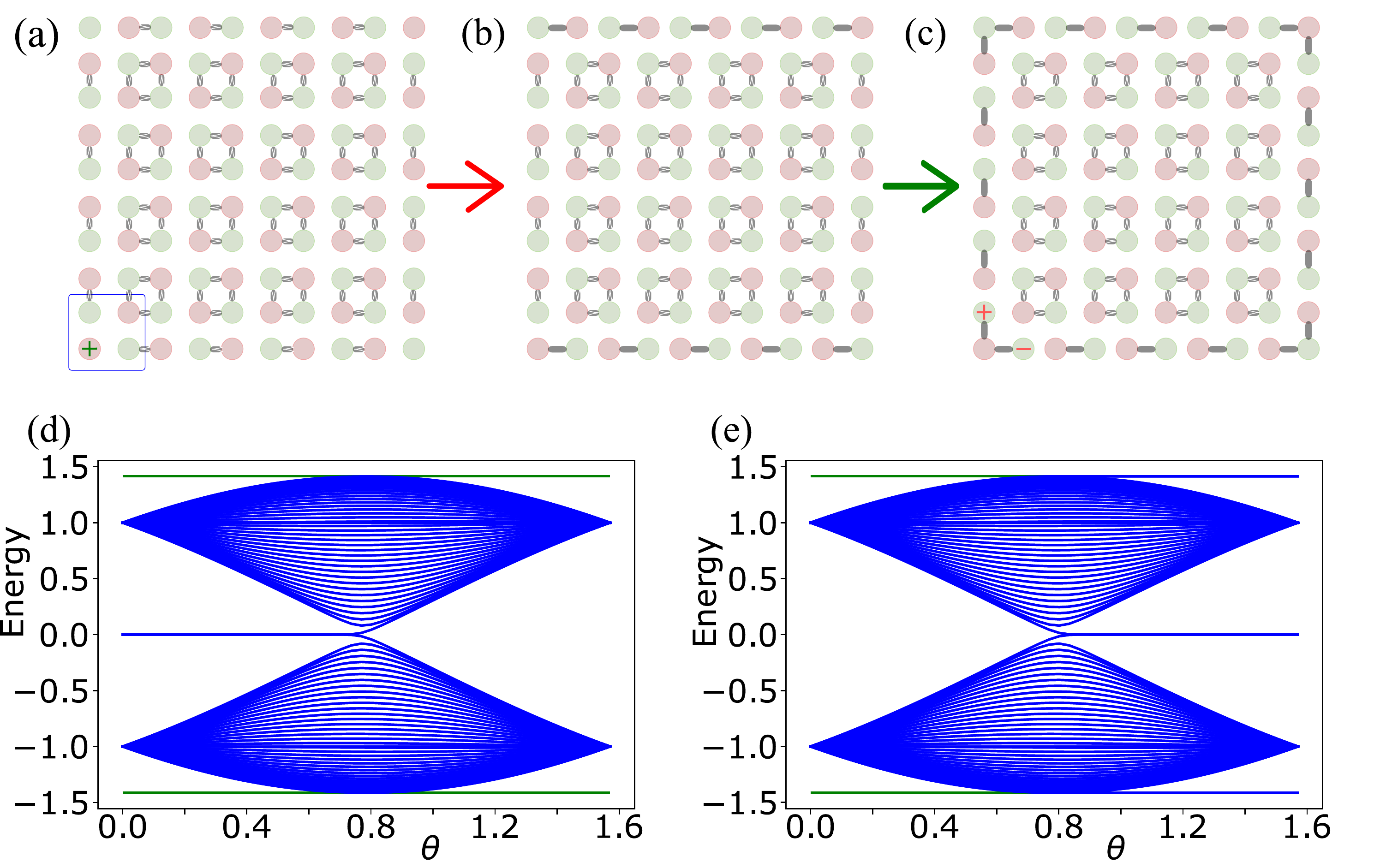}
\caption{{\bf Chiral-symmetric two-dimensional insulators feature corner modes protected by edge topology. } (a) ${\mathcal C}_4$-symmetric model. Each plaquette is threaded by a $\pi$-flux. (b) a strong ${\mathcal C}_4$-symmetric edge perturbation is applied, leaving the corner modes intact. However, before the band-closing the corner mode of the lower left corner had a chiral charge $\chi = +1$, after the  band-gap reopening its value has changed to $\chi = -1$. (c) If a strong edge perturbation is applied along the vertical edges only, thus breaking the ${\mathcal C}_4$-symmetry, we find that the corner modes do not survive the edge band-gap closing and reopening. The spectral flow for going from (a) to (b) is shown in panel (d), whereas the spectral flow in going from (b) to (c) is shown in (e). In both cases the intra- and inter-cell hopping parameters  $t_\textrm{edge}$ and $t'_\textrm{edge}$ along the perturbed edges are given by $t_\textrm{edge} = t\sin(\theta)$ and $t_\textrm{edge}' =t'\cos(\theta)$. Furthermore, each of the four edges  are $40$ unit cells long, and the bulk bands at $\pm \sqrt{2}  t'$ are shown in green. \label{fig:C4model}}
\end{figure}

The microscopic tight-binding model we will consider is schematically depicted in Fig.~\ref{fig:C4model}(a).
It possesses an internal conventional chiral symmetry and a ${\mathcal C}_4$ fourfold rotational symmetry. 
When considered in an open-disk geometry that respects both the rotational and the chiral symmetries, the system features four zero-energy states which are completely localized at the corners of the lattice when the intra-cell hopping amplitude $t\equiv 0$. 
Precisely as the end modes of the SSH chain, these corner modes can be characterized with a  $\mathbb{Z}$ number corresponding to their chiral charge $\chi$. 
More importantly, we find that the the corner modes remain pinned at zero energy under the influence of perturbations that preserve the chiral symmetry, and neither close the edge nor the bulk band gap. 
This follows from the same reasoning which is behind the stability of zero energy modes in the SSH atomic chain. In particular, it should be stressed that this stability does not rely on the fourfold rotational symmetry. 

There is, however, an important distinction between the edge states of the SSH chain and the corner states of a two-dimensional chiral-symmetric insulator. The presence of edge states in the SSH atomic chain is only dependent on the topology of the one-dimensional bulk Hamiltonian. On the contrary, the presence of a corner mode in a two-dimensional chiral-symmetric insulator is dependent on both the topology of the neighboring edges and the topology of the two-dimensional bulk. To illustrate this point, we apply to the model shown in Fig.~\ref{fig:C4model}(a) an edge perturbation 
that can be strong enough to close and reopen the edge gap. 
 Specifically, we introduce a perturbation on the top and bottom edges [c.f. Fig.~\ref{fig:C4model}(b)], thereby explicitly breaking the fourfold rotational symmetry. The spectral flow obtained by increasing the strength of the edge perturbation [c.f. Fig.~\ref{fig:C4model}(d)] shows that the zero-energy modes remain pinned at zero energy in the weak perturbation regime. However, in the strong edge perturbation regime, {\it i.e.} after the closing and reopening of the edge band gap, the corner modes disappear. 
 Importantly, the edge perturbation leaves the bulk of the crystal completely intact independent of its strength. 
This shows that
zero-energy modes carry a halfway protection: they are removable only by edge perturbations causing closing and reopening of the edge band gap.  
In the language of Ref.~\cite{kha19}, the chiral-symmetric insulating model shown in Fig.~\ref{fig:C4model}(a), therefore represents a boundary-obstructed topological phase protected by the chiral symmetry. 
We note that  in these phases
it is impossible to disentangle the edge topology from the bulk topology. Therefore, the presence of a zero-energy corner state by itself does not provide insights in the bulk topology alone.

Even though the chiral symmetry on its own is insufficient to stabilize the corner modes against strong edge perturbations, we find that the 
additional presence of the 
fourfold rotational symmetry does offer this kind of protection. 
To show this, we consider an additional edge perturbation along the left and right edge, increasing the strength of which eventually leads to the configuration shown in Fig.~\ref{fig:C4model}(c) where the ${\mathcal C}_4$ symmetry has been restored. The ensuing spectral flow shown in Fig.~\ref{fig:C4model}(e) shows that an additional edge gap closing and reopening point leads to a revival of the zero-energy modes. Therefore, 
the fourfold symmetric system is characterized by zero energy modes independent of the presence of a strong edge perturbation, qualifying them as the manifestation of  the two-dimensional bulk topology. In other words, supplementing the chiral symmetry with the $\mathcal{C}_4$ symmetry turns the model into a second-order topological insulator. This can be also seen using the following argument: in the concomitant presence of the chiral and ${\mathcal C}_4$ rotational symmetry, 
each corner mode
is at the intersection of two adjoined edges related to each other by the fourfold rotational symmetry. The chiral charge $\chi$ of this corner mode will change to 
$\chi\rightarrow\chi +j$, with $j$ an integer upon closing an reopening the band gap along one of the two edges. On the other hand, owing to the fourfold 
rotational  symmetry, 
the same band gap closing and reopening will occur on the other edge, which will thus contribute with an additional and equal change of the chiral charge. Hence, in total we find that the chiral charge of the corner mode 
is modified as $\chi\rightarrow \chi +2 \times j$. 
This therefore implies that, by virtue of the ${\mathcal C}_4$ symmetry, the parity of the chiral charge, and consequently the parity of the number of zero modes per corner $\nu$, is invariant under strong edge perturbations and thus represents a proper ${\mathbb Z}_2$-invariant of such a two-dimensional insulator.
 Note that this topology is considerably weaker than the topology of the SSH chain, as the latter is characterized by a $\mathbb{Z}$-number and does not necessitate the presence of a rotational symmetry.

\subsection*{Corner charge-mode correspondence} 
We now show that the topological immunity of the zero-energy boundary modes in chiral-symmetric insulators with a fourfold rotational symmetry can be also proved using a corner charge-mode correspondence that can be generalized to chiral insulators with an even-fold rotational symmetry. We first recall that, as shown in Refs.~\cite{mie18,ben19}, the crystalline topological indices characterizing rotational symmetric two-dimensional insulators are revealed in the fractional part of the corner charge.
Specifically, for corners whose boundaries cross at a maximal Wyckoff position with a site symmetry group that contains the $n$-fold rotational symmetry, the fractional part of the corner charge is a topological $\mathbb{Z}_n$ number  uniquely determined by the symmetry labels of the occupied Bloch states at the high-symmetry momenta 
in the Brillouin zone.

\begin{figure}
\includegraphics[width=1\columnwidth]{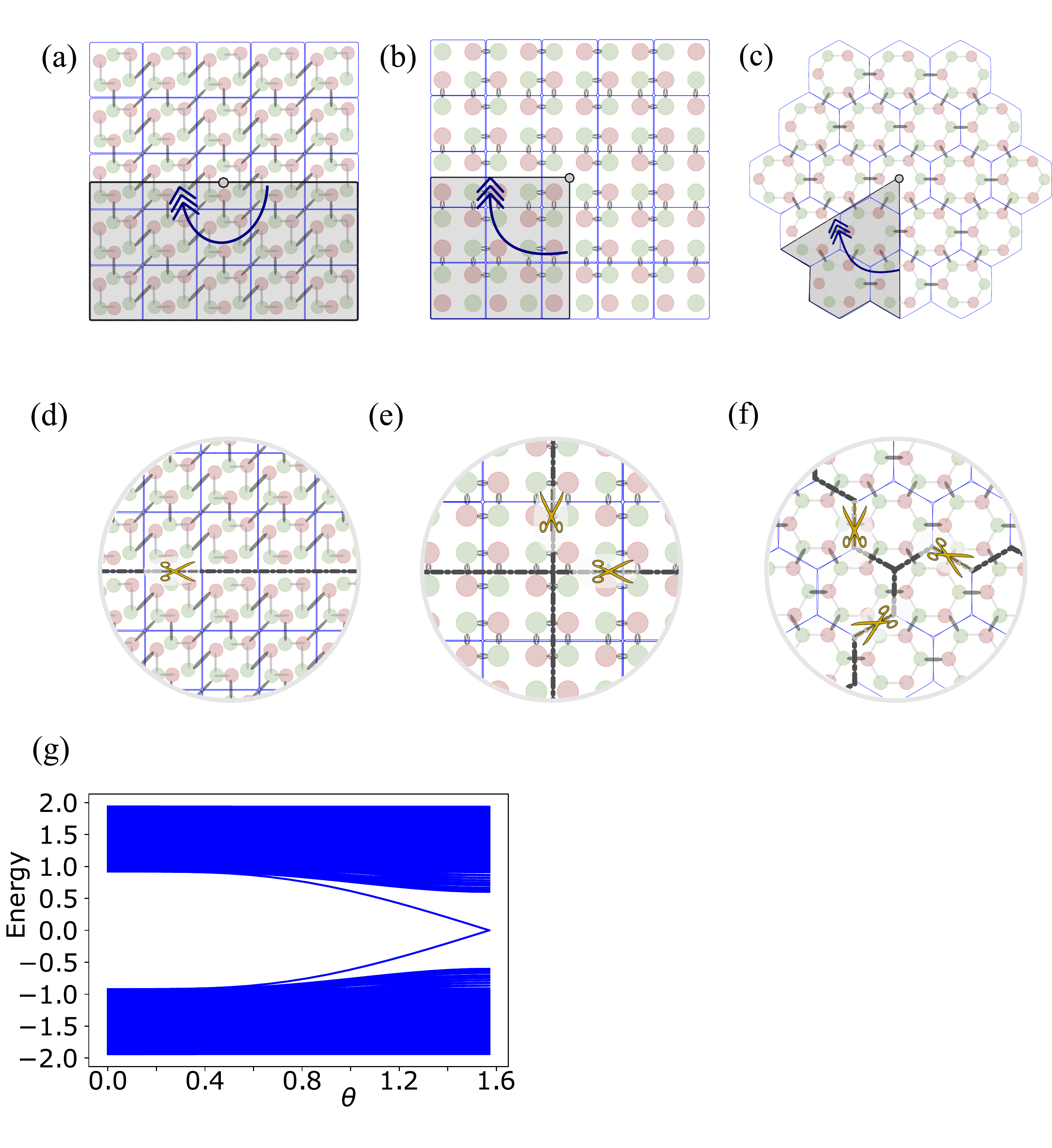}
\caption{{\bf Rotation symmetric lattices and their spectral flow by adiabatic cutting procedure.} (a)-(c) $\mathcal{C}_n$-symmetric lattices, with rotational-symmetric corner charge regions in gray that are congruent with the rotational-symmetric unit-cell center. (d)-(f) cutting lines to go from closed to open boundary conditions. (g) Spectral flow for cutting the $\mathcal{C}_4$-symmetric model of Fig.~4 open. Here, we have used $t=0.3 t'$, and the inter-cell hoppings crossing the cutting line are equal to $t'\cos(\theta)$. Each of the four edges is $20$ unit cells longs. Please note that the in-gap modes are doubly degenerate, i.e. in total there are four in-gap states. \label{fig:bulkcorner}}
\end{figure}

Generally speaking, the correspondence between these bulk topological indices and the fractional corner charge is not reflected in the presence or absence of corned modes. However, we now show that a direction relation exist between the fractional part of the corner charge and the parity of zero-energy corner modes in chiral-symmetric insulators featuring a two-, four-, or six-fold rotational symmetry. We will in fact demonstrate that the parity of the number of zero-energy states per corner $\nu$ obeys the formula 
\begin{align}
\nu& = \left( 2 \times Q_{v.b.} + N_\textrm{total}/n \right) \textrm{ modulo }2.
\label{eq:chargemode}
\end{align}
In the equation above, $N_\textrm{total}$ is the total number of sites per unit cell, $n$, as before, the order of the rotational symmetry, and  $Q_{v.b.}$ the corner charge due to the valence bands, which is a quantity quantized to $0$ or $1/2$ modulo $1$. Before deriving the relation above, a few remarks are in order. First, 
we wish to emphasize that the corner charge $Q_\textrm{v.b.}$ should be computed as the total charge within a $\mathcal{C}_n$-symmetric corner region that is congruent with unit cell centers whose site symmetry group contains the $\mathcal{C}_n$ rotational symmetry. This implies that the boundaries of the corner region should be related to each other via the rotational symmetry, and that the boundaries of the region cross at the $\mathcal{C}_n$-symmetric unit cell center. 
Example of these corners for $\mathcal{C}_2$, $\mathcal{C}_4$, and $\mathcal{C}_6$ symmetric system are shown in gray in Figs.~\ref{fig:bulkcorner}(a)-(c). Second, we remark that the direct relation between corner charge and corner modes is strictly valid for geometric configurations in which the finite-size system can be tiled using an integer number of unit cells. This exclude the presence of fractional unit-cells in the open-disk geometry. Note that the lattice structures shown in Figs.~\ref{fig:bulkcorner}(a)-(c) obey this constraint. Finally, we will consider lattices where:
 {\it i}) the positions in the unit cell with a site symmetry group containing the $\mathcal{C}_n$ symmetry  do not host any atomic site and {\it ii}) atomic sites do not intersect the unit-cell boundaries.
Note that these conditions automatically ensures that the total number of atomic sites per unit cell is a multiple of the rotational symmetry order $n$. 

We can now derive the corner charge-mode correspondence of Eq.~\ref{eq:chargemode}, and first note that at full filling the corner charge is equal to $N_\textrm{total}/n$ modulo $2$. By inspection of Fig.~\ref{fig:bulkcorner}, it can be easily seen that in all three lattice structures $N_\textrm{total}/n = 1$. 
Next, we note that the corner charge at full-filling can be decomposed into three separate contributions: a valence band $Q_{v.b.}$, a conduction band $Q_{c.b.}$, and finally an in-gap state $Q_\textrm{in-gap}$ contribution. 
Here, the valence (conduction) band part accounts for all states whose energies $E$ lie below (above) the band gap, i.e. $E \leq \Delta/2$ ($E\geq \Delta/2$), with $\Delta$ the band gap. The in-gap contribution instead is due to states whose energies $E$ lie inside the band gap, i.e., $|E|<\Delta/2$. 
Furthermore, the presence of chiral symmetry guarantees that the corner charge due to the conduction band is identically equal to the corner charge due to the valence band, i.e. $Q_{v.b.}=Q_{c.b.}$. Note that the latter is a true equality of number, which includes also the integer part of the charge. 
As a result, we obtain the following:
\begin{align*}
N_\textrm{total}/n& = Q_{v.b.} + Q_{c.b.} +Q_\textrm{in-gap} \textrm{ modulo }2\\
&= 2 \times Q_{v.b.} + Q_\textrm{in-gap} \textrm{ modulo }2.
\end{align*}
Finally, we use that the in-gap contribution $Q_\textrm{in-gap}$ is equal, modulo $2$, to the parity of the number of corner states per corner $\nu$. Hence, upon rearranging terms we arrive at Eq.~\ref{eq:chargemode} that proves the correspondence between fractional corner charge and the presence of zero-energy corner modes. As the corner charge is a direct probe of the two-dimensional bulk topology~\cite{mie18,ben19}, Eq.~ \ref{eq:chargemode} implies that the parity of the the number of zero-energy states per corner $\nu$ itself is a manifestation of this bulk topology. 

In principle, one may repeat the above analysis for $\mathcal{C}_3$-symmetric insulators that are also chiral-symmetric. However, an interesting interplay between the chiral symmetry and the three-fold rotational symmetry unfolds. Namely, the chiral symmetry requires that at half-filling the corner charge is a multiple of $1/2$, instead the threefold rotation symmetry implies that the corner charge is a multiple of $1/3$. Satisfying both conditions leaves only one option, namely a vanishing corner charge. In other words, the chiral symmetry renders the corner charge trivial in a $\mathcal{C}_3$-symmetric insulator, in the same way that time-reversal symmetry renders the Chern number trivial.  Therefore, we find that the mere presence or absence of zero-energy corner states in a $\mathcal{C}_3$-symmetric insulator does not shine any new light on the two-dimensional bulk topology. In fact, any  $\mathcal{C}_3$-symmetric configuration that can be tiled with an integer number of unit-cells will fail to exhibit an odd number of zero-energy states per corner. To prove this, let us suppose that each of the three corners would host a single zero-energy state. This would then imply that the lattice as a whole exhibits a chiral imbalance. However, this is at odds with the original assumption that the finite geometry can be tiled with an integer number of unit-cells. Ergo, $\mathcal{C}_3$-symmetric insulators fail to exhibit an odd number of zero-energy modes per corner.

The corner charge-mode correspondence shows that in $\mathcal{C}_2$, $\mathcal{C}_4$, and $\mathcal{C}_6$-symmetric insulators the presence of in-gap zero modes can be regarded as a particularly simple probe of bulk topology as long as the chiral symmetry is present. Put differently, the presence of in-gap zero modes serves as a proxy for the corner charge in chiral-symmetric insulators. In systems where the chiral symmetry is broken such a direct relation does not hold any longer. This, however, does not completely exclude the possibility to probe the bulk crystalline topology using spectral informations because of the presence of the so-called  \textit{filling anomaly} \cite{ben19}. The filling anomaly can be understood by noticing that for a ${\mathcal C}_n$-symmetric crystal in an open-disk geometry -- as in Fig.~\ref{fig:bulkcorner} we consider  finite size crystals that are rotation-symmetric around the unit cell center --  we can relate the corner charge to the total number of electrons modulo $n$ via $\# \textrm{electrons}  = n \times  Q_{v.b.}$. On the other hand, an identical crystal with periodic boundary conditions would host $N_F \textrm{ modulo } n$ electrons, as the central unit-cell contributes $N_F$ electrons, whereas the surrounding ones come in $n$-fold multiples of $N_F$. The filling anomaly arises when the total number of electrons for open-disk geometry does not match with the number of electrons of the corresponding lattice with periodic boundary conditions, {\it i.e.} 
\begin{align*}
n \times Q_{v.b.} &\neq N_F \textrm{ modulo } n.
\end{align*}
This also implies that the mismatch between the open and closed lattices is encoded in the difference $\delta =  n\times Q_{v.b.}-N_F$. We next perform a {\it gedankenexperiment} where we assume to continuously interpolate between the periodic and open-disk boundary conditions. Specifically, we imagine to start with periodic boundary conditions, and then cut open the two-dimensional torus ${\mathbb T}_2$ as shown in Fig.~\ref{fig:bulkcorner} (d)-(f). This can be achieved by continuously tuning to zero the amplitudes for hoppings crossing the black cutting lines in Fig.~\ref{fig:bulkcorner}(d)-(f). 
During this process $\delta$ states will cross the Fermi level $E_F$ either from below or above (depending on the sign of $\delta$). 
Indeed, this is the only way to resolve the filling anomaly. For the $\mathcal{C}_4$-symmetric model studied above, we find that upon tuning the system from a torus to an open-disk precisely two modes cross the Fermi level, see Fig.~\ref{fig:bulkcorner}(g). Note that in the absence of a chiral symmetry there is nothing that prevents the in-gap states crossing $E_F$ from dissolving into the bulk valence or conduction bands at the end point of the spectral folow. In other words, to probe the crystalline topology of a rotation-symmetric crystal via spectral methods one should track the evolution of the spectrum throughout the entire opening of the torus ${\mathbb T}_2$ . While the described cutting procedure may seem rather complex, we envision that it can be implemented for artificially created lattices. We would like to emphasize that the cutting procedure can also be used away from half-filling. Hence, the spectral flow is able to convey more information regarding the crystalline topology than the detection of an in-gap zero mode does. Finally, we note that also this spectral tool cannot be used to probe the crystalline topology of $\mathcal{C}_3$-symmetric lattices for the very simple reason that a triangular geometry cannot be obtained from opening a torus. Thus, for $\mathcal{C}_3$-symmetric insulators there is no spectral signature of the bulk crystalline topology, and one can only resort to corner charge probes.

\section{Discussion}
To sum up, we have shown that precisely as boundary modes in atomic chains, corner modes are a generic feature of two-dimensional band insulators and are not necessarily a signature of topology. 
Contrary to previous theoretical and experimental claims, we have indeed proved that the corner modes encountered the breathing kagome lattice do not
exhibit any kind of topological robustness,
and have to be instead qualified as ordinary boundary modes.
We have also contrasted the fragility of these modes with the robustness of the zero-energy modes appearing in insulators equipped with an internal chiral symmetry. 
When taken alone, the chiral symmetry provides the 
halfway topological robustness characteristic of the recently introduced boundary-obstructed topological phases. Furthermore, the  presence of a rotational symmetry provides an additional protection mechanism that qualifies the zero energy modes as the prime physical consequence of a higher-order bulk topology. 

We have also proved that the immunity of the topological corner modes in the concomitant presence of rotational and chiral symmetry directly follows from a one-to-one correspondence between fractional corner charges, which reveal the crystalline topology of generic insulators, and the parity of the number of zero modes. This one-to-one correspondence only works in crystals possessing an even-fold rotational symmetry, and thus exclude ${\mathcal C}_3$-symmetric crystals such as the kagome lattice.  We wish to remark, however, that even though the breathing kagome lattice does not display a genuine bulk-corner correspondences, its underlying crystalline topology is still reflected in the fractional charge at corners or other topological defects, such as dislocations~\cite{mie18r}.

In closing, we would like to highlight that the topological immunity of zero energy modes in insulators is very different in nature from the one provided by time-reversal symmetry in conventional" first-order topological insulators. This is immediately apparent from the fact that the helical edge states of a two-dimensional topological insulator represent anomalous states: Their anomaly reside in the fact that it is impossible to find a one-dimensional insulator with an odd number of Kramer pairs at the Fermi energy. 
On the contrary, the zero-energy fermionic modes encountered in chiral-symmetric insulators do not represent an essential anomaly for the very simple reason that a quantum dot with an odd number of fermionic modes is entirely allowed in nature. 
This difference is not only interesting {\it per se}. In fact, a topological boundary mode that is also anomalous carries an additional degree of protection. Consider for instance a two-dimensional topological insulator whose edges are brought in close proximity to a time-reversal invariant nanowire. In such combined system the edge boundary modes will survive, at odds with what would happen in a Su-Schrieffer-Heeger atomic chain if the termination is changed by an atomic site addition. Moreover, one can also proceed in the opposite direction. For example, one may modify a two-dimensional insulator that is both $\mathcal{C}_3$-symmetric and chiral-symmetric but fails to exhibit any in-gap mode by attaching single sites to its three corners. Even though this would result in the presence of robust zero-energy corner modes, it should be stressed that these modes are nothing but a microscopic detail of the corner and edges. Put differently, these modes are not in any way informative of the bulk topology.

\begin{appendix}

\section{Topological zero modes in chiral symmetric atomic chains}
The robustness of the zero modes in a chiral-symmetric atomic chain can be rigorously proved using the following:
 let $H(\lambda)$ parametrize the 
chiral-symmetric perturbed Hamiltonian (with $\lambda$ the perturbation strength) of a semi-infinite atomic chain that features a single left 
end state $|\Psi_L(0)\rangle$ at zero energy. 
To show that 
this edge state will remain pinned at zero
energy, we can 
track the left edge state $|\Psi_L(\lambda)\rangle$, assuming $\lambda$ is switched on adiabatically.
The chiral symmetry guarantees that every state $|\chi\rangle$ at an energy $+E\neq 0$ has a chiral partner $\sigma_z|\chi\rangle$ at energy $-E$. Hence, end states can only move away from zero energy in pairs. Thus, 
we can conclude that the parity of the number of left edge states pinned at zero energy is robust against all continuous perturbations that respect the chiral symmetry. 
Note that although this conclusion is strictly valid for a semi-infinite chain, it also holds for finite chains, {\it i.e.} with a left and a right edge, as long as
the decay length of the edge states is small compared to the system size. In particular, 
this argument cannot be used
when the perturbation closes the bulk band gap, 
since in the latter case the decay length of the boundary states always reaches the system's size. This additionally shows
that the presence of the zero-energy edge states is 
an intrinsic feature of the one-dimensional bulk Hamiltonian.

For completeness, we finally note that, being a bulk quantity,  the $\mathbb Z$ chiral charge $\chi_L$ of the topological zero modes 
can be expressed as a winding number in terms of the one-dimensional Bloch Hamiltonian 
$$H(k) = \begin{pmatrix}
0&h(k)\\
h^\dagger(k)&0
\end{pmatrix}$$ using the following formula:
\begin{align*}
\chi_L&=\frac{i}{2\pi}\int_0^{2\pi}\mathrm{d}k \frac{d}{d k}\log{\det{h(k)}}.
\end{align*}
For a detailed derivation of the above bulk-boundary correspondence, we refer the reader to Ref.~\onlinecite{ryu02}.

\end{appendix}

\end{document}